\documentclass{article}
\usepackage{graphicx}
\usepackage{latexsym}

\begin{document}
\begin{flushright}
UM-P 048-2000\\
RCHEP 011-2000
\end{flushright}

\begin{center}
{\bf\huge Radion signature in $\gamma\gamma$ scattering}\\
\hspace{10pt}\\
S.R.Choudhury\footnote{src@ducos.ernet.in} \\
{\em Department of Physics, Delhi University, Delhi, India},\\
A.S.Cornell\footnote{a.cornell@tauon.ph.unimelb.edu.au}
and G.C.Joshi\footnote{joshi@physics.unimelb.edu.au}\\
{\em School of Physics, University of Melbourne,}\\
{\em Parkville, Victoria 3108, Australia}\\
\hspace{10pt}\\
$5^{th}$ of December, 2000
\end{center}
\hspace{10pt}\\
\begin{abstract}
\indent We investigate the phenomenological consequences of the
possible existence of the radion at a TeV-scale for $\gamma\gamma$ 
scattering in the TeV range.  We find that polarized
cross-sections only give possible experimental signatures for
the radion.
\end{abstract}

\indent Theories with extra spatial dimensions, where the standard model (SM) 
fields are localized on a brane, have of late been the subject of many 
investigations \cite{anton}.  These are typically Kaluza-Klein type theories, but
have features distinctively of their own.  Of these models, the one proposed by
Randall and Sundrum \cite{rs} is a particularly simple one.  This is based on
a 5-dimensional space-time with a non-factorizable metric.  The extra dimension
is a single $S^1/Z_2$ orbifold wherin there are two branes of opposite tensions, 
one physical and the other hidden from us, at the two boundary points of the
orbifold.  There is also a bulk cosmological constant, and with a proper choice
of this constant and the brane tensions, RS are able to solve the 5-dimensional
Einstein equations to get a metric in vacuum of the form \\
\begin{equation}
\label{one}
ds^2=e^{-2kr_c|\phi|}\eta_{\mu\nu}dx^{\mu}dx^{\nu}-r_c^2d\phi^2
\end{equation}
\indent In equation (1), $k$ is a parameter comparable to the
5-dimensional Planck mass $M$, $r_c$ is the radius of the compact dimension
represented by $\phi$.  RS are able to show that the four dimensional Planck
scale $M_{pl}$ is related to the above parameters by
\begin{equation}
\label{two}
M_{pl}^2= \frac{M^3}{k}(1-e^{-2kr_c\pi})
\end{equation}
so that with $k \approx M$, $M_{pl}$ is of the order of $M$.  However, because
of the exponential factor in the metric, fields confined to the brane have
mass scales $M e^{-k r_c\pi}$, so that with $kr_c\approx 12$ , the weak scale
is generated from the Planck scale, thus solving the hierarchy problem.\\
\indent The metric, equation (1), is the background value of the more
general form
\begin{equation}
\label{three}
ds^2=e^{-2k|\phi|T(x)}g_{\mu\nu}dx^{\mu}dx^{\nu}-T^2(x)d\phi^2
\end{equation}
where $g_{\mu\nu}$ is the 4-d gravitational field and $T(x)$ is the radion
field.  Goldberger and Wise \cite{gold} showed that the radion field
fluctuations are a massless scalar field that couples to matter with a scale set
by the TeV-scale.  As presented in \cite{anton}, no mechanism was given for
stabilizing the radion field at its background value.  A model which does
that was later given by Goldberger and Wise \cite{gold} involving a bulk scalar
field propagating in the background metric.  They were able to show that the
value of $kr_c\approx12$ needed to solve the hierarchy problem leads to a
massive scalar radion fluctuation field with TeV-scale mass that is most
likely to be lighter than the KK-excitations of the graviton.\\
\indent Phenomenologically, one can expect the radion-SM field coupling, roughly
in the TeV-range, to be the first candidate to produce experimental signals
should these underlying ideas have validity \cite{gold,csaki}.\\
\indent Phenomenological signals for possible radion mediated interactions are of
course most likely to be detected through small changes in the precision
electroweak parameters like the W-Z mass difference or the weak mixing angle
\cite{mah,cheung}. However, processes where the SM values are suppressed for
some reason are also good testing grounds for these kind of non SM interactions.
$\gamma\gamma$ scattering, a process which can be experimentally realized
at the NLC using high-energy $\gamma$ beams obtained by backward Compton
scattering of laser photons off $e^+,e^-$ beams, is one such process since
in the SM, the process has no tree level contribution and is thus of order
$\alpha^2$ coming through box graphs involving W, Z and fermi particles.
In this note, we consider the phenomenology of radion signals in TeV-scale
$\gamma\gamma$ scattering.\\
\indent The SM analysis of $\gamma\gamma$ scattering has been done first by Jikia
and Tkabladze \cite{jikia} and later by Gounaris, Poryfyradis and Renard
\cite{gou1,gou2}.  We denote the invariant amplitude for the process
\begin{eqnarray}
\gamma(p_1,\lambda_1)+\gamma(p_2,\lambda_2)
\rightarrow\gamma(p_3,\lambda_3)+
\gamma(p_4,\lambda_4)
\end{eqnarray}
as $F(\lambda_1,\lambda_2,\lambda_3,\lambda_4;s,t,u)$ 
where s,t,u are the Mandelstam variables.  Neglecting terms of the order of
$M_w^2/s$, $M_w^2/t$, $M_w^2/u$, the non-negligible SM contributions are:
\begin{eqnarray}
\label{four}
F(++++;s,t,u)/\alpha^2 & = & 
12 - 12\left(\frac{u-t}{s}\right) 
\left[\ln\left(\frac{-u-i\epsilon}{M_w^2}\right) - 
\ln\left(\frac{-t-i\epsilon}{M_w^2}\right)\right] \nonumber \\
& & - 16\left(1-\frac{3tu}{4s^2}\right)\left[\left\{
\ln\left(\frac{-u-i\epsilon}{M_w^2}\right) -
\ln\left(\frac{-t-i\epsilon}{M_w^2}\right)\right\}^2 +\pi^2\right] \nonumber \\
& &+16s^2\left[\ln\left(\frac{-s-i\epsilon}{M_w^2}\right)
\ln\left(\frac{-t-i\epsilon}{M_w^2}\right)/\left(st\right)\right] \nonumber \\
& &+16s^2\left[\ln\left(\frac{-s-i\epsilon}{M_w^2}\right)
\ln\left(\frac{-u-i\epsilon}{M_w^2}\right)/\left(su\right)\right] \nonumber\\
& &+16s^2\left[\ln\left(\frac{-t-i\epsilon)}{M_w^2}\right)
\ln\left(\frac{-u-i\epsilon}
{M_w^2}\right)/\left(tu\right)\right] \\
& & -8-8\left(\frac{u-t}{s}\right)
\left[\ln\left(\frac{-u-i\epsilon}{M_w^2}\right) -
\ln\left(\frac{-t-i\epsilon}{M_w^2}\right)\right] \nonumber \\
& & -4\left(\frac{t^2+u^2}{s}\right) 
\left[\left\{\ln\left(\frac{-u-i\epsilon}{M_w^2}\right) 
- \ln\left(\frac{-t-i\epsilon}{M_w^2}\right)\right\}^2+\pi^2\right] \nonumber
\end{eqnarray}
\begin{eqnarray}
\label{five}
F(+++-;stu)/\alpha^2 & = & F(++--;stu)/\alpha^2 \approx-12+8Q^4
\end{eqnarray}
where $Q^4$ is the sum over the fourth power of charges of all the fermions
in SM.  The radion contribution to the amplitudes is dictated by the interaction
\begin{equation}
\label{six}
L_{int} = \frac{1}{\Lambda}\cdot T^{\mu}_{\mu}\cdot \phi,
\end{equation}
where $\Lambda$ is the vev of the radion field $\phi$, and $T$ is the
energy momentum tensor.  For massless photons, there is as such no
direct  $\gamma\gamma\phi$ coupling, but that arises because of the
trace anomaly and also (in the same order of $\alpha$) through triangle
diagrams involving fermions.  These work out to be effective $\gamma\gamma
\phi$ vertex given approximately by \cite{cheung}:
\begin{eqnarray}
\lambda \left( p_1.p_2 g_{\mu\nu} - p_{2\mu}p_{1\nu}\right)
\end{eqnarray}
where
\begin{eqnarray}
\lambda=\left(\frac{i\alpha}{2\pi\Lambda}\right) \cdot \left[
\frac{19}{6} - \frac{41}{6} - 2\right]
\end{eqnarray}
\indent With the vertex given as above, the resonating radion contribution is
non-vanishing for $F(++++;stu)$, $F(++--;stu)$ and their parity images.  Denoting
the radion contribution by a superscript $R$, we get
\begin{equation}
\label{seven}
F^R(++++;stu)=F^R(++--;stu)
\end{equation}
\begin{equation}
\label{eight}
F^R(++++;stu)=\frac{4\lambda^2s^2}{s-m^2+i m\Gamma_{tot}}
\end{equation}
where $\Gamma_{tot}$ is the total decay width of the radion, which we assume is
to the WW and ZZ channels and given by:
\begin{equation}
\label{nine}
\Gamma_{tot}=\frac{3m^3}{32\pi<\phi>^2}
\end{equation}
The total helicity amplitudes is the sum of the SM and radion
contribution and these give the total unpolarized cross-section
given by
\begin{equation}
\label{ten}
\frac{d\sigma_0}{d(\cos\theta)} =
\frac{1}{128s}\left[|F(++++)|^2+|F(+++-)|^2+|F(++--)|^2\right]
\end{equation}
If we chose a typical value of the radion mass $m$ and the vev $<\phi>$ as\\
$m=<\phi>=0.6$ Tev, the effect of the radion on the unpolarized cross-section is
less than $1\%$ as shown in figure (1).  It perhaps will be a very difficult
signature to be detected experimentally through measurement of unpolarized
cross-sections.\\
\begin{figure}
\includegraphics[angle=270,width=10cm]{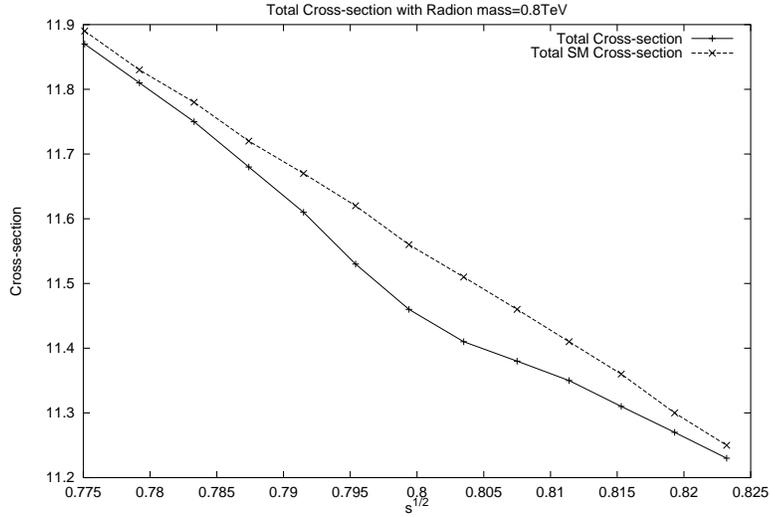}
\caption{A comparison of the total cross-sections of
$\gamma\gamma\to\gamma\gamma$
for the case of just the SM and also with the inclusion of the radion field, with
radion mass of $0.8TeV$}
\label{fig1}
\end{figure}
\begin{figure}
\includegraphics[angle=270,width=10cm]{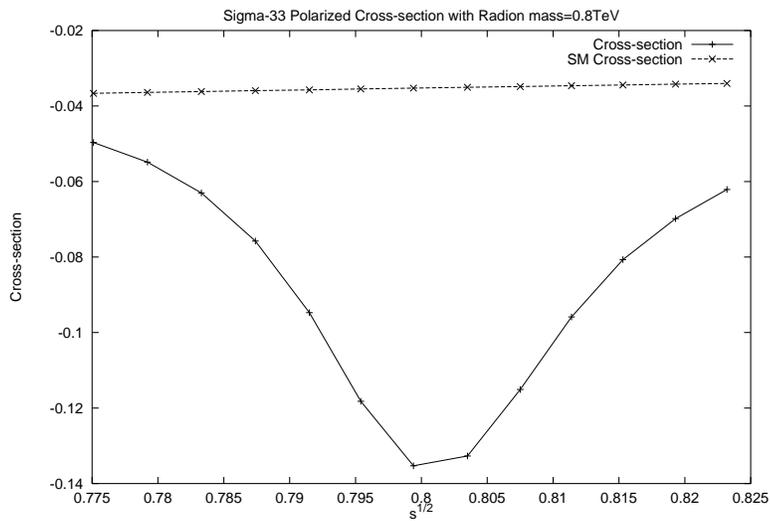}
\caption{A comparison of the sigma-33 cross-sections of
$\gamma\gamma\to\gamma\gamma$
for the case of just the SM and also with the inclusion of the radion field, with
radion mass of $0.8TeV$}
\label{fig2}
\end{figure}
\indent The radion contribution given above is numerically small but is unique
in its helicity structure.  Both the SM loops and possible contributions
from spin-2 KK excitations of the graviton, do not have any resonating
s-channel contributions to the amplitudes $F(++++;stu)$ and $F(++--;stu)$.  This
implies that in polarized cross-sections, the radion signature will
be more pronounced and distinctive in at least some of them.  Indeed,
using the polarized cross-sections given in \cite{gou1}:
\begin{eqnarray}
\label{eleven}
\frac{d\sigma}{d(\cos\theta)} & = & \frac{d\sigma_0}{d(\cos\theta)} +
<\xi_2\bar{\xi}_{2}>\frac{d\bar{\sigma}_{22}}{d(\cos\theta)} 
 + \left[<\xi_3>\cos(2\phi)+ <\bar{\xi_3}>\cos(2\bar{\phi})\right]
\frac{d\bar{\sigma}_{3}}{d(\cos\theta)} \nonumber \\
& & + <\xi_3\bar{\xi}_{3}>\left[\frac{d\bar{\sigma}_{33}}{d(\cos\theta)} 
 \cos(2(\phi+\bar{\phi})) + \frac{d\bar{\sigma}_{33}'}{d(
\cos\theta}\cos(2(\phi-\bar{\phi})) \right] \nonumber \\
& & + \left[<\xi_2\bar{\xi_3}> \sin(2\bar{\phi})- <\xi_3\bar{\xi_2}>
\sin(2\phi)\right] \frac{d\sigma_{23}}{d(\cos\theta)}
\end{eqnarray}
where the $\xi$'s are the Stokes parameters.  Of particular interest in
obtaining a signature of the radion contribution arises from the
quantity $\frac{d\bar{\sigma}_{33}'}{d(\cos\theta)}$, given in terms of the
helicity amplitudes as:
\begin{equation}
\label{twelve}
\frac{d\bar{\sigma}_{33}}{d(\cos\theta)}=\frac{1}{128s}\sum_{\lambda_3\lambda_4}
Re[F(++\lambda_3\lambda_4)F^*(--\lambda_3\lambda_4)]
\end{equation}
We show in figure (2), estimates of the cross-section in equation (15)
integrated over the region $30^o<\theta< 150^o$ as a function of
energy.  It is clear that around the resonance, there will be a very
visible dip in this cross-section.  This feature is very specific to
to the spin zero radion since a spin-2 KK-excitation of the graviton
do not show such a phenomenon to this cross-section \cite{src1,src2}.\\
\indent In conclusion, we find that a signature of a typical radion resonance
with masses and a vev in the TeV range is possible in a polarized
cross-section, but is rather insignificant in an unpolarized cross-section 
for $\gamma\gamma$ scattering.

\end{document}